\documentclass{article}
\usepackage{url}
\usepackage{longtable}
\usepackage{float}
\usepackage{subfloat}
\usepackage{lscape,epsfig}
\usepackage{graphicx}
\usepackage{amssymb}
\usepackage{amsmath}
\usepackage{natbib}
\textheight=22cm \DeclareSymbolFont{ppa}{OT1}{ppl}{m}{it}
\DeclareMathSymbol{\vv}{\mathalpha}{ppa}{'166}

minus 2.3mu \thickmuskip = 2.6mu plus 2mu minus 2.6mu

\let\svthefootnote\thefootnote

\begin{document}

\newcommand{\dd}{\,{\rm d}}
\newcommand{\ie}{{\it i.e.},\,}
\newcommand{\etal}{{\it $et$ $al$.\ }}
\newcommand{\eg}{{\it e.g.},\,}
\newcommand{\cf}{{\it cf.\ }}
\newcommand{\vs}{{\it vs.\ }}
\newcommand{\zdot}{\makebox[0pt][l]{.}}
\newcommand{\up}[1]{\ifmmode^{\rm #1}\else$^{\rm #1}$\fi}
\newcommand{\dn}[1]{\ifmmode_{\rm #1}\else$_{\rm #1}$\fi}
\newcommand{\upd}{\up{d}}
\newcommand{\uph}{\up{h}}
\newcommand{\upm}{\up{m}}
\newcommand{\ups}{\up{s}}
\newcommand{\arcd}{\ifmmode^{\circ}\else$^{\circ}$\fi}
\newcommand{\arcm}{\ifmmode{'}\else$'$\fi}
\newcommand{\arcs}{\ifmmode{''}\else$''$\fi}
\newcommand{\MS}{{\rm M}\ifmmode_{\odot}\else$_{\odot}$\fi}
\newcommand{\RS}{{\rm R}\ifmmode_{\odot}\else$_{\odot}$\fi}
\newcommand{\LS}{{\rm L}\ifmmode_{\odot}\else$_{\odot}$\fi}

\newcommand{\Abstract}[2]{{\footnotesize\begin{center}ABSTRACT\end{center}
\vspace{1mm}\par#1\par \noindent {~}{\it #2}}}

\newcommand{\TabCap}[2]{\begin{center}\parbox[t]{#1}{\begin{center}
  \small {\spaceskip 2pt plus 1pt minus 1pt T a b l e}
  \refstepcounter{table}\thetable \\[2mm]
  \footnotesize #2 \end{center}}\end{center}}

\newcommand{\TableSep}[2]{\begin{table}[p]\vspace{#1}
\TabCap{#2}\end{table}}

\newcommand{\FigCap}[1]{\footnotesize\par\noindent Fig.\  %
  \refstepcounter{figure}\thefigure. #1\par}

\newcommand{\TableFont}{\footnotesize}
\newcommand{\TableFontIt}{\ttit}
\newcommand{\SetTableFont}[1]{\renewcommand{\TableFont}{#1}}
\newcommand{\MakeTable}[4]{\begin{table}[htb]\TabCap{#2}{#3}
  \begin{center} \TableFont \begin{tabular}{#1} #4
  \end{tabular}\end{center}\end{table}}

\newcommand{\MakeTableSep}[4]{\begin{table}[p]\TabCap{#2}{#3}
  \begin{center} \TableFont \begin{tabular}{#1} #4
  \end{tabular}\end{center}\end{table}}

\newenvironment{references}%
{ \footnotesize \frenchspacing
\renewcommand{\thesection}{}
\renewcommand{\in}{{\rm in }}
\renewcommand{\AA}{Astron.\ Astrophys.}
\newcommand{\AAS}{Astron.~Astrophys.~Suppl.~Ser.}
\newcommand{\ApJ}{Astrophys.\ J.}
\newcommand{\ApJS}{Astrophys.\ J.~Suppl.~Ser.}
\newcommand{\ApJL}{Astrophys.\ J.~Letters}
\newcommand{\AJ}{Astron.\ J.}
\newcommand{\IBVS}{IBVS}
\newcommand{\PASP}{P.A.S.P.}
\newcommand{\Acta}{Acta Astron.}
\newcommand{\MNRAS}{MNRAS}
\renewcommand{\and}{{\rm and }}
\section{{\rm REFERENCES}}
\sloppy \hyphenpenalty10000
\begin{list}{}{\leftmargin1cm\listparindent-1cm
\itemindent\listparindent\parsep0pt\itemsep0pt}}%
{\end{list}\vspace{2mm}}

\def\TYLDA{~}
\newlength{\DW}
\settowidth{\DW}{0}
\newcommand{\dw}{\hspace{\DW}}

\newcommand{\refitem}[5]{\item[]{#1} #2%
\def\REFARG{#3}\ifx\REFARG\TYLDA\else, {\it#3}\fi
\def\REFARG{#4}\ifx\REFARG\TYLDA\else, {\bf#4}\fi
\def\REFARG{#5}\ifx\REFARG\TYLDA\else, {#5}\fi.}

\newcommand{\Section}[1]{\section{#1}}
\newcommand{\Subsection}[1]{\subsection{#1}}
\newcommand{\Acknow}[1]{\par\vspace{5mm}{\bf Acknowledgments.} #1}
\pagestyle{myheadings}

\newfont{\bb}{ptmbi8t at 12pt}
\newcommand{\xrule}{\rule{0pt}{2.5ex}}
\newcommand{\xxrule}{\rule[-1.8ex]{0pt}{4.5ex}}

\begin{center}
{\Large\bf
 The Cluster AgeS Experiment (CASE).\dag  \\
 Variable stars in the field of  
 the globular cluster M10}{\LARGE$^\ast$}
 \vskip1cm
  {\large
      ~~M.~~R~o~z~y~c~z~k~a$^1$,
      ~~W.~~N~a~r~l~o~c~h$^{1,2,3}$,
      ~~A.~~S~c~h~w~a~r~z~e~n~b~e~r~g~~--~~C~z~e~r~n~y$^1$
      ~~I.~B.~~T~h~o~m~p~s~o~n$^4$
      ~~R.~~P~o~l~e~s~k~i$^5$
      ~~and~~W.~~P~y~c~h$^1$,
   }
  \vskip3mm
{ $^1$Nicolaus Copernicus Astronomical Center, ul. Bartycka 18, 00--716 Warsaw, Poland\\
     e--mail: (mnr, wnarloch, pych, alex)@camk.edu.pl\\
  $^2$ Universidad de Concepci\'on, Departamento de Astronomia, Casilla 160-C, Concepci\'on, Chile\\
  $^3$ Millennium Institute of Astrophysics, Santiago, Chile\\
  $^4$The Observatories of the Carnegie Institution for Science, 813 Santa Barbara
      Street, Pasadena, CA 91101, USA\\
     e--mail: ian@obs.carnegiescience.edu\\
  $^5$ Department of Astronomy, Ohio State University, 140W. 18th Ave., Columbus, OH43210, USA\\
     e--mail: poleski.1@osu.edu\\
}
\end{center}

\vspace*{7pt}
\Abstract 
{The field of the globular cluster M10 (NGC 6254) was monitored between 1998 and 2015 in a search 
for variable stars. $V$-light curves were derived for 40 variables or likely variables,
most of which are new detections. Proper motions obtained within the CASE project indicate that 
18 newly detected variables and 14 previously known ones are members or likely members of the 
cluster, including one RRc-type, three type II Cepheids, and 14 SX~Phe-type pulsators, one contact 
binary, and six semi-regular red giants. As a byproduct of the search we discovered a candidate 
binary comprised of main sequence stars with the record-short orbital period of 0.042 d. We also confirmed the 
photometric variability of the red straggler M10-VLA1 hinted at by Shishkovsky et al. (2018), who discovered 
this object spectroscopically. In Appendix 1 we show that CASE proper motion measurements are in a good 
agreement with those retrieved from the {\it Gaia} archive, while Appendix 2 presents evidence for 
low frequency $\gamma$ Doradus-type oscillations in SX Phe stars belonging to M10. 
}
{globular clusters: individual (M10) -- stars: variables -- stars: SX Phe -- blue stragglers -- stars:
individual (M10-VLA1)
}

\let\thefootnote\relax\footnotetext{\dag CASE was initiated and for long time led
by our friend and tutor Janusz Kaluzny, who prematurely passed away in March 2015.}
\let\thefootnote\relax\footnotetext
{$^{\mathrm{\ast}}$Based on data obtained with the Swope telescope at  
Las Campanas Observatory.}
\let\thefootnote\svthefootnote

\Section{Introduction} 
\label{sec:intro}
M10 (NGC 6254) is projected against the outskirts of the Galactic bulge at $l=15\zdot\arcd1$, 
$b=23\zdot\arcd1$, in an appreciably reddened region with the total $E(B-V)$ reddening varying between 
0.26 mag and 0.29 mag across our field of view\footnote{The extinction calculator at \url{
https://irsa.ipac.caltech.edu/applications/DUST/} was used for this estimate.}. Its core radius 
$r_c$, half--mass radius $r_h$, tidal radius $r_t$, [Fe/H] index, radial velocity, heliocentric 
distance $d_\odot$, and galactocentric distance $d_G$ are equal to 0\zdot\arcm86, 1\zdot\arcm81, 
21\zdot\arcm6, -1.52, 75.8$\pm$1.0 km s$^{-1}$, 4.3 kpc and 4.6 kpc, respectively 
(Harris 1996, 2010 edition; hereafter H10).\footnote{Webpage 
\url{http://vizier.u-strasbg.fr/viz-bin/VizieR?-source=VII/202}.} Among globular clusters (GCs), 
M10 is distinguished by an almost purely blue horizontal branch (Dotter et al. 2010), and
a very low specific frequency of RR Lyr-type variables: $S_{RR}=1.1$, where
\begin{displaymath}
S_{RR}=N_{RR}10^{0.4(7.5+M_{Vt})},
\end{displaymath}
$N_{RR}$ is the number of RR Lyrs, and $M_{Vt}$ is the integrated $V$-band luminosity of the 
cluster (H10). The mass of M10 and its present relaxation time at $r_h$ are estimated to be 
1.55$\times10^5 M_\odot$, and 743 Myr, respectively (Webb et al. 2017). The latter authors find 
the degree of mass segregation and the global mass function of the cluster to be consistent with 
its dynamical age, which makes M10 to be the only GC with a well understood dynamical history in 
the sample they study.  

Even though M10 is as close to the Sun as 47 Tuc (H10), it has been much less extensively explored. 
Pre--CCD searches for variables, summarized by Clement et al. (2001, 2017 edition\footnote {Webpage 
\url{http://www.astro.utoronto.ca/~cclement/cat/C1654m040}.}; hereafter C17), resulted in the 
detection of just four variable objects. Within the targeted CCD surveys performed so far (von Braun 
et al. 2002, Salinas et al. 2016; hereafter S16) additional 15 variables were found in the cluster 
field, including three clear nonmembers. Blue straggler stars (BSS) in M10 were investigated by 
Dalessandro et al. (2011, 2013). Those authors identified 120 candidate BSS, 
however the time-coverage of their data was insufficient for an accompanying variability study. 
Pietrukowicz et al. (2008) searched the cluster for dwarf novae, but none was found. 
Finally, a radio-bright red straggler suspected of photometric variability has recently been 
discovered in M10 by Shishkovsky et al. (2018; hereafter S18).

Our survey is a part of the CASE project (Kaluzny et al. 2005) conducted using telescopes of the Las 
Campanas Observatory, and its aim is to increase the inventory of variable objects in the field of M10. 
In Section~2 we briefly report on the observations, explain the methods used to calibrate the 
photometry, and briefly introduce methods employed to identify variable stars. Newly discovered 
variables are presented and 
discussed in Section~3, whereas Section~4 contains new data on previously known variables which we 
consider worthy of publishing. For all the variables the membership probability is given based of 
proper motion measurements of Narloch et al. (2017; hereafter N17). The paper is summarized in Section~5,
and in Appendix 1 proper motions of N17 are compared to those retrieved from the $Gaia$ archive. 
\section{Observations and data processing}

\label{sec:obs}
The present paper is mainly based on images acquired on the 1.0--m Swope telescope 
equipped with the $2048\times3150$ pixel SITe3 camera which provided a field of view 
$14.8\times 22.8$ arcmin$^2$ at a scale of 0.435 arcsec/pixel. The data were 
collected during two seasons, 1998 and 2002, comprising 32 nights between 
May 1998 and June 2002. Additional observations were performed during seven 
nights starting from June 28, 2015. A new E2V camera was  used, with the same scale 
of 0.435 arcsec/pixel, and with a field of view subrastered from the original 
$29.7\times29.8$ arcmin$^2$ to that of SITe3. The same set of filters was always used.
 The seeing ranged from 1\zdot\arcs2  to 3\zdot\arcs2 and 1\zdot\arcs2  
to 2\zdot\arcs5 for $V$ and $B$ frames, respectively, with median values of 1\zdot\arcs4 
and 1\zdot\arcs5. For the analysis, 1207 $V$--band images and 161 $B$--band images were 
selected. 

The photometry was performed using an image subtraction technique implemented in the 
DIAPL package.\footnote{Available from http://users.camk.edu.pl/pych/DIAPL} 
To reduce the effects of PSF variability, each frame was divided into 4$\times$6  
overlapping subframes. The reference frames were constructed by combining 11 images 
in $V$ and 4 in $B$ with an average seeing of 1\zdot\arcs22 and 1\zdot\arcs23, respectively.
The light curves derived with DIAPL were converted from differential counts to magnitudes 
based on profile photometry and aperture corrections determined separately for each 
subframe of the reference frames. To extract the 
profile photometry from reference images and to derive aperture corrections, the 
standard Daophot, Allstar and Daogrow (Stetson 1987, 1990) programs were used. 
Profile photometry was also extracted for each individual image, enabling useful 
photometric measurements of stars which were overexposed on the reference frames. 
\subsection{Photometric calibration and search for variability}

\begin{figure}[H]
   \centerline{\includegraphics[width=0.95\textwidth,
               bb = 27 317 560 560, clip]{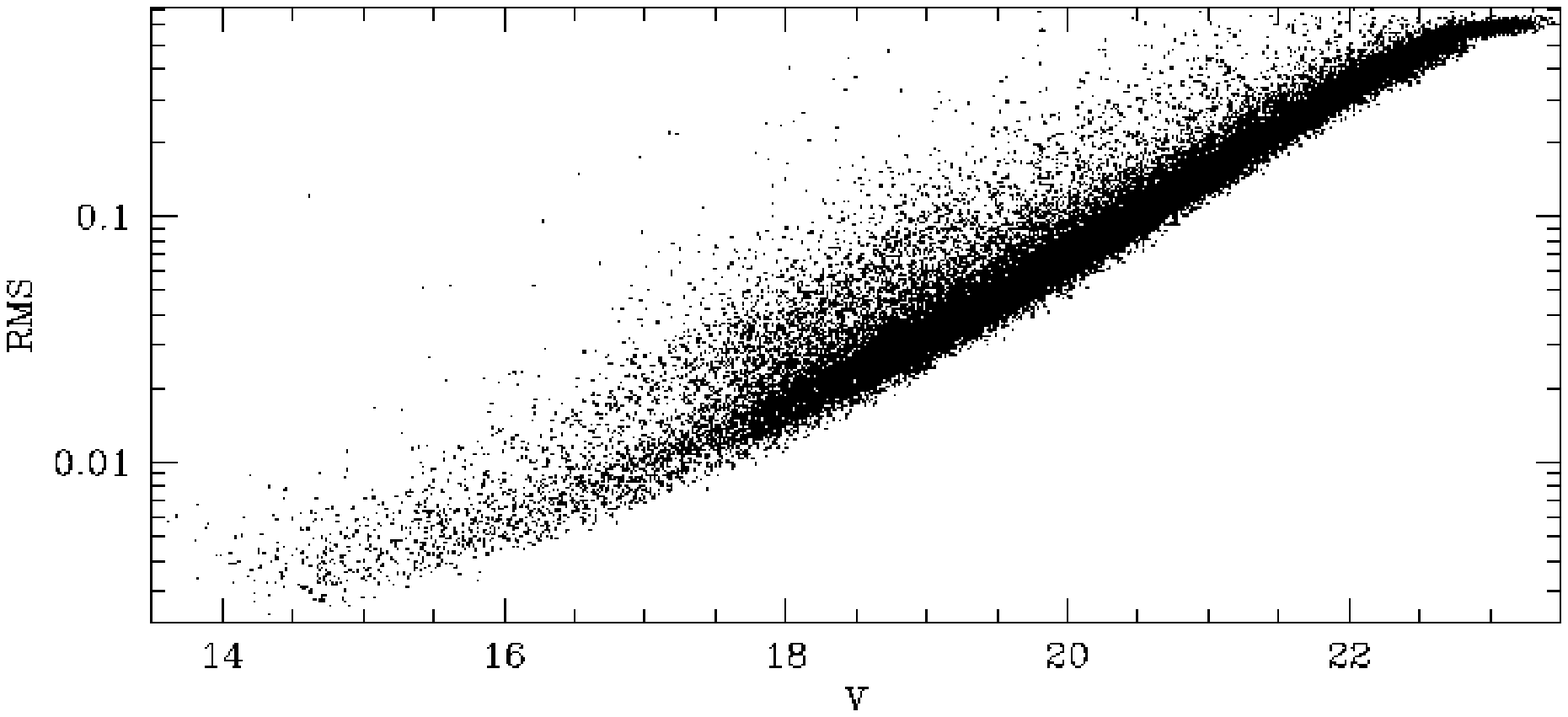}}
   \caption{ Standard deviation vs. average $V$-band magnitude for
    light curves of stars from the M10
 field. 
    \label{fig:rms}}
\end{figure}

The calibration of SITe3 data was based on observations of 24 Landolt 
standards, yielding the following transformation to the standard system:
\begin{align}
  V &= v + 2.9236(49) + 0.0071(70)\times(b-v) \nonumber\\
  B - V &= 0.2349(40) + 1.0438(55)\times(b-v) \nonumber,
\end{align}
where lower case and capital letters denote instrumental and standard magnitudes, 
respectively, and numbers in parentheses are uncertainties of the last significant 
digits. The standard SITe3 magnitudes were then used to transform the instrumental 
E2V values. Since M10 has a relatively loosely populated central part, we were able 
to reach a photometric accuracy of 0.1 mag at $V=21$ mag (Fig. \ref{fig:rms}). 

We obtained time--series photometry for 45,942 stars brighter than $V\sim$22 mag, and 
conducted a search for periodic variables using the AOV and AOVTRANS algorithms 
implemented in the TATRY code (Schwarzenberg--Czerny 1996 and 2012;  
Schwar\-zenberg--Czerny \& Beaulieu 2006). 

\section{Variable stars and their membership in M10} 
\label{sec:vars}

Membership of the cluster was assigned based on i) proper motions (PM) measured by N17, ii) angular 
distances from the center of the cluster, and iii) CMD locations combined with the variability type.
Details concerninng PM measurements and calculations of membership class $C_{PM}$ 
and membership probability $P_{PM}$ are given in N17, who also provide a PM catalog for 
nearly 450000 stars in the fields of 12 GCs. As detailed Appendix 1, their PMs of M10 
variables generally agree with those of $Gaia$, discrepant values being obtained for two 
objects only.
We consider a variable to be a member or likely member of the cluster if one of the following 
criteria is fulfilled: 
\begin{enumerate}
\item $P_{PM}\geq$70\%.
\item $P_{PM}<$70\%, $C_{PM}=1$ or 2, CMD-location compatible with cluster membership, variability 
      type compatible with CMD--location, and geometric membership probability 
      $P_{geom}=1-\pi r^2/S>90$\%, where $r$ is star's angular distance 
      from the center of M10 ($\alpha$ = 16$^{\mathrm h}$ 57$^{\mathrm m}$ 
      09\zdot$^{\mathrm s}$05, $\delta$ = -04\arcd 06\arcm 01\zdot\arcs1) in arcseconds, and 
      $S=1.22\times10^6$ is the size of the field of view in arcseconds$^2$.
\item Proper motion not known, but $P_{geom}>70$\%, CMD-location compatible with cluster 
      membership, and variability type compatible with CMD--location.
\end{enumerate}

Light curves were obtained for all the known variables within our field of view, and for 
24 new variable or likely variable stars, 18 of which are PM  or likely PM--members of M10. 
Membership status was also assigned to the variables previously discovered.\footnote{Data 
for all the identified variables are available at http://case.camk.edu.pl} 

\begin{figure}[H]
   \centerline{\includegraphics[width=0.95\textwidth,
               bb = 53 198 560 557, clip]{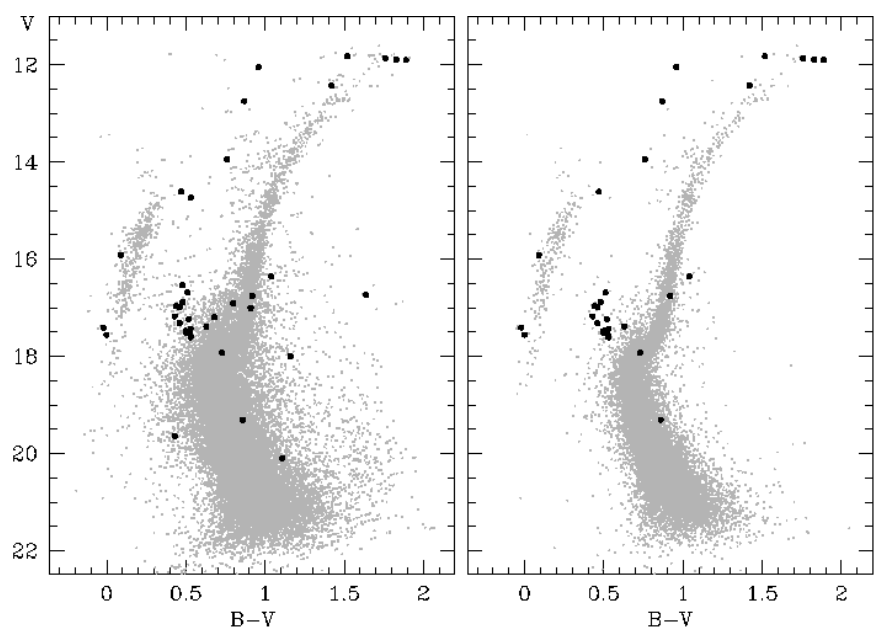}}
   \caption{CMD for the observed field. Left: all stars 
    for which proper motions were measured by N17. Black points mark all the variables detected 
    within the present survey for which $B$-band magnitudes were available. Right: same 
    as in the left panel, but for PM-members of the cluster only. 
    \label{fig:cmds}}
\end{figure}

The color-magnitude diagram
(CMD) of the observed field, constructed based on the reference images, is shown in
Fig.~\ref{fig:cmds}. To make it readable, stars with proper motions measured by N17 were 
only selected to serve as a background against which the variables are plotted. Stars 
identified by N17 as PM-members of the cluster are shown in the right panel.

Basic data for the  variables are given in Table \ref{tab:CASE}. For our naming 
convention to agree with that of C17 we start numbering the new variable cluster members 
from V17. The stars whose PM indicate that they do not belong to M10 are given names 
from N1 on. The equatorial coordinates for epoch J2000 are given in columns 2 and~3.
They conform to the UCAC4 system (Zacharias et al. 2013), and are accurate to 
0\zdot\arcs2 -- 0\zdot\arcs3 (statistical 1-$\sigma$ errors). The $V$--band 
magnitudes in column 4 correspond to the maximum light in the case of eclipsing binaries, 
and  the average magnitude is given for the remaining cases. Columns 5--7 give $B-V$ color, 
amplitude in the $V$--band, and period of variability. 

\begin{figure}[H]
   \centerline{\includegraphics[width=0.95\textwidth,
               bb = 40 40 418 553, clip]{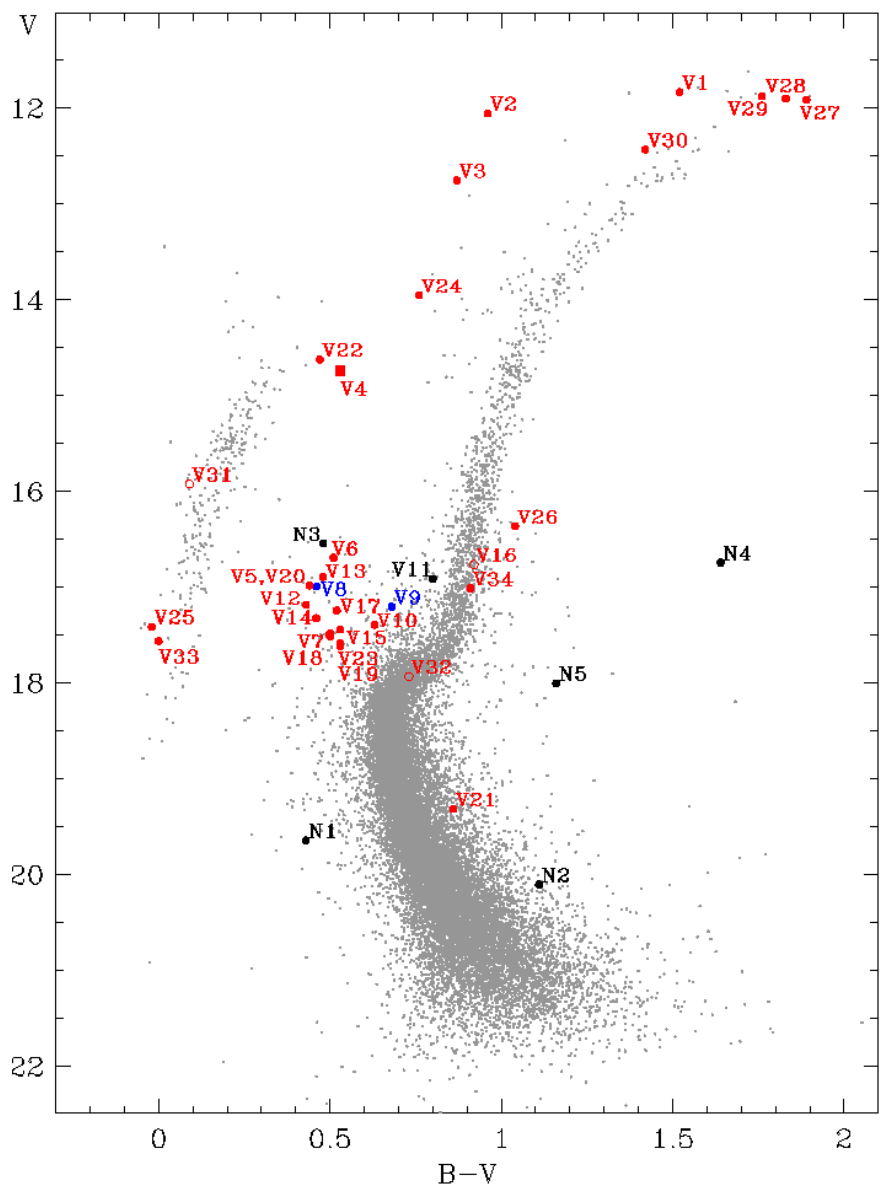}}
   \caption{CMD for the observed field with locations of the variables identified  
    within the present survey. Red: members of M10; blue: discrepant CASE and {\it Gaia}
    proper motions, black: field stars. Filled circles: confirmed variables; open circles: 
    suspected variables; square: constant star listed as a likely RR Lyr-type pulsator by C17. 
    The gray background stars are the same as in the right panel of Fig.~\ref{fig:cmds}.
    \label{fig:cmd_var}}
\end{figure}

Fig. ~\ref{fig:cmd_var} shows the CMD of M10 with identifications of variable stars. 
PM--members of the cluster are marked in red, field stars in black, and two objects 
with discrepant PMs - in blue. The gray background stars are the PM--members of M10 
from the right panel of Fig.~\ref{fig:cmds}. 
In the following, we describe the new variables whose light curves are shown in 
Figs.~\ref{fig:CASE1} and \ref{fig:CASE2}. 

\subsection{Cluster members}
\label{sec:members}
The blue stragglers V17 -- V20 are SX Phe-type pulsators. Multimodal pulsations are observed for 
V17 and V19, and are likely for V18 (in all three cases the appreciable dispersion of the light-curve 
in Fig.~\ref{fig:CASE1} is at least partly caused by or can be at least partly attributed 
to amplitude variations). V20 in turn is a showcase example of a High Amplitude 
$\delta$~Scuti-type variable with 
an amplitude of $\sim$0.5 mag and a very stable, apparently single-mode light curve. All 
four of these stars are 100\% PM-members of M10. However, V19 is located unusually far from the 
center of the cluster for a blue straggler (at $\sim4.4 r_h)$, so that in principle it 
might be a field $\delta$~Sct star. A radial velocity measurement would be needed to confirm 
its membership. 

V21, another 100\% PM-member of the cluster, is the only W UMa-type variable detected in M10.
Judging from Fig. \ref{fig:rms}, we should have easily detected W UMas brighter than $V\approx19$ 
mag  with an amplitude larger than $\sim$0.1 mag, but none were found. The apparent paucity of 
contact binaries in M10 compared to other clusters surveyed within CASE (e.g. 16 confirmed, 1 
confirmed + 7 likely, 3 confirmed + 1 likely, 7 confirmed + 1 likely, and 10 confirmed, respectively  
in M22, NGC 3201, NGC 362, NGC 6362, and M4) is puzzling. Underrepresented W UMas 
together with numerous blue stragglers, at least some of which should have originated from 
merged binaries (e.g. Li et al. 2018), suggest that - contrary to the conjecture of Webb et al. 
(2017) - M10 may have an interesting dynamical history. 
Contact systems in GCs are generally not found significantly below the main-sequence turnoff,
a possible explanation being that they form primarily due to nuclear evolution of detached
binaries, and a contact configuration is achieved once the more massive component exhausts 
hydrogen in the core and starts to expand (e.g. Kaluzny et al. 2016). V21 is only the second 
exception to this rule, the first one being KT-08 in M22 (Rozyczka et al. 2017). 

V22, the only  RR Lyr-type variable found in M10, has a light curve characteristic of first 
overtone (i.e. RRc) pulsators. It exhibits a moderate Blazhko effect, and in the CMD of the 
cluster it resides close to the blue edge of the instability strip.   

The low-amplitude sinusoidal light curve of V23 may originate from the reflection effect. Since 
the star is located in a rather loosely populated area beyond $r_h$, there should be no problem
with a spectroscopic verification of this possibility.

V24 is a BL Her-type pulsator, and the third type II Cepheid in M10 after V2 and V3 cataloged 
by C17. Among these objects it has the lowest luminosity and the shortest period ($P=2.31$ d).

The semi-regular variable V25 is located on the blue horizontal branch (BHB), which makes its 
4.46 day variability difficult to understand. Its light curve seems too irregular to be caused by the 
reflection effect. In the Swope frames there is no trace of blending, but in principle it may 
be tightly blended with a field binary. Unfortunately, no HST imaging data of M10 are available at its 
location at $\sim1.3 r_h$, and the only means to verify the nature of this object is spectroscopy. 

V26 is a red straggler with a sinusoidal light curve and possible secular changes of the average 
brightness. As such, it photometrically resembles V34 described below. The phase of the sinusoid 
is preserved throughout 1998 and 2002 seasons, and most likely until the 2015 season, which suggests 
a stable orbital origin of the variations, and makes V26 interesting for spectroscopic observations. 

PM-members V27, V28, V29 and the likely member V30 are long period semi-regular red giants with 
amplitudes of a few tenths of a magnitude. V27, V28 and V30 reside within $r_h$; V29 is
located in the 
outer part of M10 at $\sim 3.5 R_h$. In the CMD  the four objects are clustered at the red 
giant tip next to V1 which was discovered in mid-50's, and one may wonder how they escaped 
detection for over 60 years. 
 
Suspected variables V31, V32 and V33 exhibit low-amplitude, roughly sinusoidal variations of 
unknown origin which should be independently confirmed.

\begin{figure}[H]
    \centerline{\includegraphics[width=0.95\textwidth,
        bb = 45 534 563 688, clip]{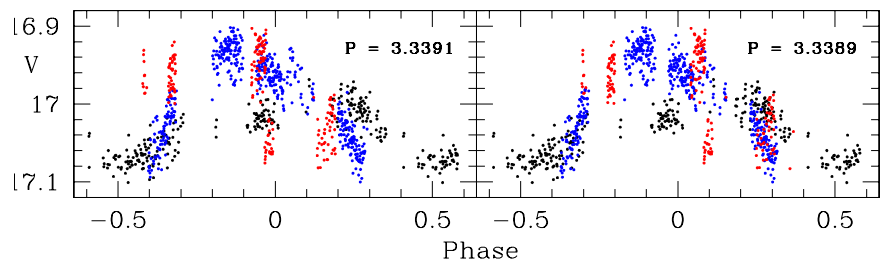}}
   \caption{$V$-band light curve of V34 phased with P=3.3391 d (left) and P=3.3389 (right).
    Black, blue and red points: seasons 1998, 2002 and 2015, respectively.
    \label{fig:v34}}
\end{figure}

V34 is a likely member of M10 and an optical counterpart of the radio- and X-ray active object
M10-VLA1 studied by S18. Throughout all three seasons it exhibits variations which can be phased 
with the spectroscopic period of 3.3391~d found by S18 (see Fig. \ref{fig:v34}; we note that 
$P=3.3389$ d seems to fit the light curve slightly better while preserving the overall agreement 
with the radial velocity data). The shape and amplitude of the light curve vary from season to season, 
thus confirming the conclusion of S18 who classify this object as an interacting binary. 
According to S18, the observed properties of M10-VLA1 are consistent with a black hole primary, 
with the exception of the low mass function which requires a statistically unlikely, nearly 
face-on orientation. Assuming that the photometric variability originates predominantly from 
the reflection effect, we tried to model the light curve with the PHOEBE implementation of the 
Wilson--Devinney code (Pr\v{s}a and Zwitter 2005), the effect itself being simulated by a hot 
spot on the surface of the secondary. We found it possible to reasonably fit the light curve even 
for an orbital inclination of 4 $\deg$. However,  the hot spot has to be placed at an orbital longitude 
of $\sim$90$\deg$, which is a physically unlikely location. Thus, either the photometric variations
originate predominantly somewhere else in the system (e.g. in an accretion disk around the primary), 
or the alternative possibility suggested by S18 is true, namely that M10-VLA1 is an extreme flaring 
RS~CVn system (note that, since we do not observe eclipses, its orbital period could be different 
from the spectroscopic period of S18). In any case, this object certainly deserves a detailed 
multi-wavelength study. 

\subsection{Field variables}
\label{sec:fieldvar}

N1 is a background  SX Phe or $\delta$ Sct-type high-amplitude pulsator discovered by von Braun et al. 
(2002) who refer to it as V3 in their paper. We confirm the period they found, and find the pulsations 
to be likely multimodal.

Based solely on its CMD-location, the contact binary N2 could in principle belong to M10. Its proper 
motion is unknown, but its large distance from the center of the cluster (almost 6$r_h$) indicates
that it is  a background object. 

N3 is an RRc-type pulsator with  possible Blazhko effect variations. The CMD-location and proper 
motion indicate it is another background object.

\begin{figure}[H]
   \centerline{\includegraphics[width=0.95\textwidth,
               bb = 42 529 528 765 clip]{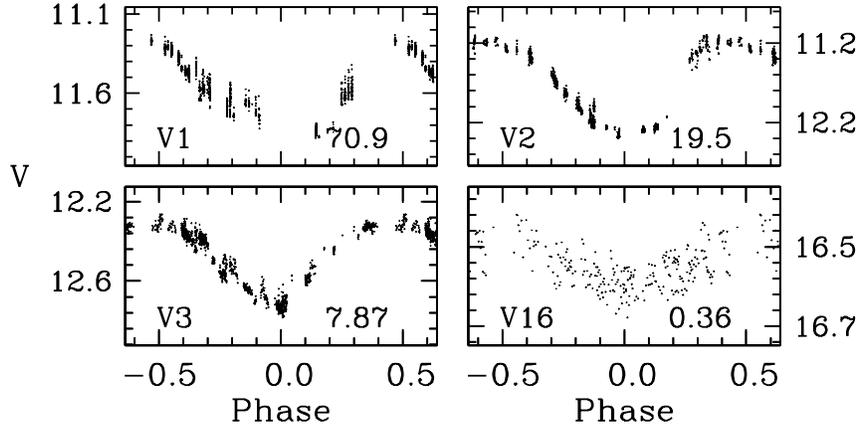}}
    \caption{Phased $V$-band light curves for selected variables from the 
     C17 catalog. Panel labels give star ID and period in days.  
     \label{fig:Clement}}
\end{figure}

The proper motions of N4 and N5 identify them as field objects. N5 is most probably a background 
red giant 
similar to V3 (see Fig. \ref{fig:Clement}), while the nature of the low-amplitude sinusoidal variability
of N4 is unclear. The variability might arise from the reflection effect, but additional data are needed to verify such a
possibility. 

N6, a marginally detected  eclipsing binary with an amplitude of $\sim0.3$ mag and a period 
of only 0.042~d (60.5 minutes), is so weak in the $B$-band that we could not detect it in 
our frames. N17 were not able to derive its proper motion, but a simple 
reasoning shows that it cannot belong to M10. If it were a member of the cluster then with 
$A_V=3\times0.28=0.84$ mag its absolute magnitude in the $V$-band would amount to 8.1 mag. The 
light curve is compatible with that generated by a pair of nearly identical stars. Assuming 
they are indeed identical, each of the components would have $M_V=8.85$ mag, corresponding to 
${\cal M}=0.60$ M$_\odot$ and $R=0.56$ R$_\odot$ (Pecaut, Mamajek \& Bubar 2012,
2018 edition\footnote{Webpage: \url{http://www.pas.rochester.edu/~emamajek/EEM_dwarf_UBVIJHK_colors_Teff.txt}}; hereafter PMB18). For this mass and $P=0.042$ d we get an orbital separation $a=0.54$ R$_\odot$ 
- much too small to accommodate two $R=0.56$ R$_\odot$ components. Since for low-mass stars $R$ scales 
roughly proportionally to $\cal{M}$, while at a fixed $P$ $a\propto{\cal M}^{1/3}$, the 
components of N6 must be much less massive than 0.60 M$_\odot$, which implies that 
the binary must be located much closer to the Sun than M10. For ${\cal M}=0.12$ M$_\odot$ 
we obtain $a=0.32$ and $R=0.15$ R$_\odot$ (PMB18), i.e. a nearly contact configuration which 
would have the observed $m_V=22.1$ mag if it was located at 340 pc from the Sun. Neglecting
absorption, this is the maximum distance allowed by our data. Still less massive configurations 
would have
to be located correspondingly closer to the Sun, down to $\sim$45 pc for ${\cal M}=0.08$ M$_\odot$. 
The shortest orbital period reported so far for a binary composed of main-sequence stars is 
0.098~d for OGLE-BLG-ECL-000066 (Soszy\'nski et al. 2015). Since N6 would
become the next record holder if its variability was confirmed, it clearly deserves
a dedicated observational effort.

\section{New data on known variables}
\label{sec:oldvar}

The red giants V1, V2 and V3 have rather stable, roughly sinusoidal light curves 
(Fig. \ref{fig:Clement}),
suggesting that their variability originates primarily from pulsations or long-living spots.

V4, cataloged by C17 as a possible RR Lyr-type variable, is in fact constant with an accuracy 
of 0.01 mag in $V$-band. As a PM-member
of M10 it is interesting because of its CMD-location in the instability strip, slightly redward
from the genuine RR Lyr-type variable V22. Since {\it Gaia's} photometry confirms our 
findings, the pair becomes an interesting 
target for a follow-up study aimed at explaining why V22 does pulsate, while V4 does not.

V5 -- V15 were identified by S16 as SX Phe-type pulsators which, with a possible exception 
of V9, belonged to M10. According to our criteria, the only nonmember of the cluster is V11,
which we tentatively classify as a foreground $\delta$~Sct-type variable. We confirm the 
very short period and complex nature of the V10 pulsations found by S16, however 
in the $V/(B-V)$ plane this star is located among the BSS rather than midway between the turnoff 
and the BHB tip as in their $i/(g-i)$ diagram. 
A similar mismatch occurs for V9 which according to S16 belongs to the BHB, whereas we find 
it in the area occupied by the BSS. An inspection of archival HST frames reveals V9 and V10 as tight 
($<$0\zdot\arcs5) blends of nearly equally bright stars. Such blended pairs are likely to have 
highly nonstandard colors, which could account for the observed discrepancies. For all the 
remaining objects from the V5 -- V15 group the CMD-locations of S16 agree reasonably well with ours. 

For V16 a steady increase in brightness of $\sim$0.02 mag was observed by S16 throughout the 
6.65~hour time span of their observations on HJD 2457224. Our data show periodic variations 
with $P=0.36$ d and an amplitude of a few hundredths of a magnitude, but in the 2015 season only 
(HJD 2457190 -- 2457201). Because of incomplete phase coverage we cannot be 100\% sure about 
their reality. We note that in 
archival HST frames V16 splits into three objects, and the suspected variability may originate 
in one of the weaker stars of the trio. 

\Section{Summary}
 \label{sec:sum}

Our photometric survey of the field of the globular cluster M10 resulted in the discovery of 24 new variable or likely 
variable stars, 18 of which are PM-members of the cluster. Cluster membership was confirmed 
for 14 out of 16 variables cataloged earlier by C17. M10 harbors a rich population of blue 
stragglers, however we did not find any eclipsing binaries among them,  unique among clusters
studied so far within CASE, i.e. NGC 6752 (Kaluzny \& Thompson 2009), M55 (Kaluzny et al. 2010),
M4 (Kaluzny et al. 2013), NGC 6362 (Kaluzny et al. 2014), M12 (Kaluzny et al. 2015), NGC 3201 
(Kaluzny et al. 2016), NGC 362 (Rozyczka et al. 2016) and M22 (Rozyczka et al. 2017). A total 
of 13 blue stragglers (nine known and four newly discovered) are identified as SX Phe-type 
pulsators, most of these are multimodal (see Appendix 2 for a detailed discussion). We stress 
that in the whole cluster just one eclipsing binary was found, suggesting a peculiar  dynamical 
history for M10. 

Stars V4 and V22 share a nearly common location in the CMD close to the edge 
of the instability strip, however only V22 is found to pulsate. A more accurate photometric 
follow-up is desirable to precisely determine their magnitudes and colors. If their location 
close to each other in the CMD is confirmed, a detailed study should be 
undertaken to explain why V4 does not vary. 

We provide a light curve of V34  -- the optical counterpart of the radio and X-ray 
source M10-VLA1 discovered spectroscopicaly by S18 -- obtained over three observing seasons, 
and we argue that the observed variability is 
unlikely to originate from the reflection effect. Another interesting object, N6, is a marginally 
detected eclipsing system with $P=0.042$d which, if confirmed, will have the record-short orbital 
period among binaries with main-sequence components. 

Finally, the three type II Cepheids V2, V3 and 
V24  represent three distinct evolutionary stages corresponding to three different crossings 
of the instability strip. 

\Acknow
{We thank Grzegorz Pojma\'nski for the lc code which vastly facilitated the work with light curves. 
ASC acknowledges partial funding from NCN grant 2016/23/B/ST9/03123.
This work is partly based on data from the European Space Agency (ESA) mission {\it Gaia} 
(\url{https://www.cosmos.esa.int/gaia}), processed by the {\it Gaia}
Data Processing and Analysis Consortium (DPAC, 
\url{https://www.cosmos.esa.int/web/gaia/dpac/consortium}). \\
Funding for the DPAC
has been provided by national institutions, in particular the institutions
participating in the {\it Gaia} Multilateral Agreement.
We also made use of the Mikulski Archive for 
Space Telescopes (MAST). STScI is operated by AURA, Inc., under NASA contract NAS5-26555. 
Support for MAST for non--HST data 
is provided by the NASA Office of Space Science via grant NNX09AF08G and by other grants and contracts.}

\section*{Appendix 1: CASE and {\it Gaia} PM-measurements}

N17 measured proper motions for 33 of the variables presented here, for 29 of which {\it Gaia} measurements are
also available. To enable a comparison of the two sets of results, the absolute {\it Gaia} PMs were transformed 
to the cluster frame iteratively. First, 
three evident nonmembers (N3, N4 and N5) were rejected, an approximate location of the 
cluster center on the $(\mu_\alpha,\mu_\delta)$ plane was found, and PMs relative to it 
were calculated. Stars with total PMs larger than 2 mas~y$^{-1}$ (V8 and V9) were then 
rejected, and the last two steps were repeated. The resulting absolute PM of M10 was 
equal to (-4.76, -6.67)~mas~y$^{-1}$, well agreeing with (-4.82, -6.18) mas~y$^{-1}$ obtained by N17. 

\begin{figure}[H]
   \centerline{\includegraphics[width=0.95\textwidth,
               bb = 42 447 563 687 clip]{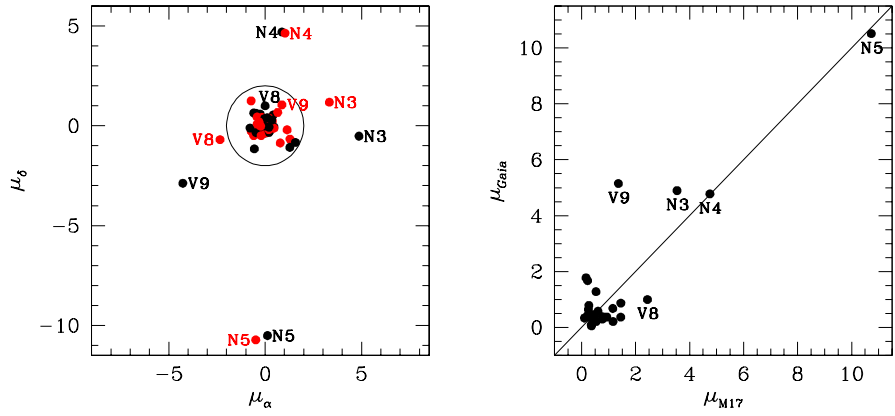}}
    \caption{Left: {\it Gaia} proper motions in M10 frame (black) compared to those of N17 (red).
Encircled are stars which, except V8 and V9, were used to derive the absolute proper motion of 
the cluster. Right: 
comparison of total proper motions $\mu = \sqrt{\mu_\alpha^2+\mu_\delta^2}$. PM unit is mas y$^{-1}$.
     \label{fig:PM_comp}}
\end{figure}

A graphical comparison of proper motions derived by N17 with those of {\it Gaia} is presented in 
Fig.~\ref{fig:PM_comp}. For N4 and N5,
which move most rapidly with respect to M10, the quantitative agreement is excellent, and 
for the slightly slower N3 it is still satisfactory. For further 24 objects a good 
qualitative agreement is observed in the sense that all of them occupy the same small area 
on the $(\mu_\alpha,\mu_\delta)$ plane, marked in Fig.~\ref{fig:PM_comp} with a 
circle. In the last two cases (V8 and V9) N17 and {\it Gaia} PMs are discrepant. This may be 
due to the fact that (as we mentioned in Section \ref{sec:oldvar}) V9 is a tight blend of 
nearly equally bright stars, whereas V8 is located at $\sim$1.2~arcsec from a star several 
times brighter.
The median distance between CASE and $Gaia$ points, taken for all objects shown in 
Fig.~\ref{fig:PM_comp}, amounts to 0.75 mas. The corresponding mean distance is 1.10 mas
(0.78 when calculated without discrepant PMs). 

RMS of $\mu_\alpha$ calculated from N17 and {\it Gaia} data for objects within the circle in
Fig.~\ref{fig:PM_comp} amounts to 0.60 and 0.54 mas y$^{-1}$, respectively, whereas RMS of $\mu_\delta$ 
- to 0.44 and 0.47 mas y$^{-1}$, respectively. At a distance of 4.3 kpc from the Sun 1~mas~y$^{-1}$ translates 
into $\sim$20 km s$^{-1}$, corresponding to an RMS of $\sim$10~km~s$^{-1}$ in $\mu_\alpha$ or $\mu_\delta$. 
As the central radial velocity dispersion of M10 is about 5~km~s$^{-1}$ (N17 and references therein), 
it is clear that while N17 results and presently available {\it Gaia} data suffice for the membership 
assignment, they are not accurate enough for a study of the internal dynamics of the cluster. An accuracy 
necessary to that end is only expected to be reached in the final $Gaia$ releases (Pancino et al. 2017).

\section*{Appendix 2: Variability of SX Phenicis stars in M10}

%
SX Phoenicis pulsating stars differ from their cousins, the $\delta$ Scuti variables
by having shorter periods, lower metalicity and smaller amplitudes. They belong to 
population II, most are members of globular clusters, galactic halo and thick disk. 
There were suspicions that many very low amplitude oscillations in them remain 
undetected (Kaluzny, 2000, Mazur et al. 2003). Space observations confirmed that belief 
(Nemec et al. 2017, hereafter NB17). 
Since some $\delta$ Scuti variables reveal low frequency oscillations of $\gamma$ Doradus
type, becoming hybrid pulsators, a similar behavior may be expected for SX Phenicis stars.
Again, recent space observations confimed this expectations (Guo et al. 2017, NB17). 
$\gamma$ Doradus stars are high-order gravity (g) mode pulsators, permitting probing of 
their interiors by means of astroseismic analysis. Since in globular clusters the SX Phe 
stars belong to the population of blue stragglers, the detection of g-modes would provide
new means to study those exotic objects. This prompted us to look more closely at 13 SX Phe 
stars discovered so far in M10. 

\subsection{Methods of Analysis}
In analysing our sample of SX Phe pulsators  one must keep in mind the peculiar form of our 
window function. In some years our observing runs spanned several months, so that a typical 
half width of 1 cycle/day (c/d) aliases is 0.01 c/d (HWHI). Nightly coverage was fairly good, 
so that  for strong, isolated peaks there was no 1 c/d ambiguity. However, due to window 
function interference this does not hold for low amplitude peaks in dense spectral regions. 
The observations typically spanned around 1500 d (6300 d for V5) with no coverage on some 
years, so that no unique cycle count was possible. As a result, daily aliases split into 
$\sim$~5 (or more for V5) yearly aliases of comparable height. Because of that we refrain 
from a discussion of mode combination and rotational splitting. Frequency and amplitude of 
the strongest peak are provided as they appear, ignoring yearly aliases of comparable height.

For the analysis the NFIT software package written by ASC was used. First, we calculated 
Analysis of Variance (AOV)
frequency spectrum, and identified its strongest peaks, next we fitted the corresponding Fourier 
series and subtracted it from observational date to pre-whiten these frequencies. In the process 
we adjusted both the amplitudes and the frequencies of sine/cosine modes by a non-linear least 
squares procedure (NLSQ). Then the whole procedure was repeated till no significant features 
remained in the spectrum. At each step the results were inspected to identify the largest harmonics 
and/or strongly correlated modes. Finally, we performed the NLSQ fit of the original data with 
the grand-total model of all detected modes. The results are listed in Table \ref{t10}. 
The median standard deviation of the residuals from model fits is 11 mmag. 

\subsection{Results}

We list only modes with amplitudes five times larger than their LSQ errors (5$\sigma$ criterion). 
For such a purpose the AOV frequency spectrum was particularly useful as it yields model to noise 
power ratio (Schwarzenberg-Czerny 1996). A common magnitude cut-off proved impractical, as due to 
the strong modulation of amplitudes and/or frequencies of some modes their errors varied considerably.
Table \ref{t10} lists formal errors of frequencies and amplitudes returned from NLSQ procedure. 
Frequency errors assume a fixed cycle count and ignore year aliasing, and amplitude errors may 
suffer from underestimation due to correlation of residuals (Schwarzenberg-Czerny 1991). Thus,
the corresponding entries in Table \ref{t10} should not be taken at their face values, but rather 
as indicators of relative fit quality of different modes. 

Two frequency regions, namely the low frequency one, below 5 c/d, and the high frequency one above this
limit deserve a separate discussion. On the one hand, because of our sampling pattern nothing reliable 
may be said on frequencies near 0 and 1 c/d ($\pm0.02$ c/d). In a few oscillations with frequencies 
lower than 1.5 c/d the poor phase coverage manifested itself by large (>0.9) correlation with the 
Fourier constant term, yet this was not common. On the other hand, at least in a half of our sample 
pre-whitening of several discrete low frequency modes left no power excess near zero frequency. Hence, 
we infer that instrumental effects did not significantly distort the power spectrum there. This is 
strengthened by the fact that no low frequency oscillations were detected in V12. The low-frequency 
part of the spectrum may be affected by variable blending effects; yet, given the fraction of variable 
stars of $\sim0.01$ in the observed field, this may not apply generally. 

At higher frequencies two alternate situations emerge. In most frequency spectra one or several well 
separated peaks and their harmonics appear in the range above 10 c/d. On rare occasions in the region 
10-30 c/d we observe one or two broad bumps several c/d wide, while little power is seen at other 
frequencies. Since FWHI of our daily frequency patterns is of order of 0.02 c/d, several dozens of 
densely packed discrete modes would be needed to produce such a feature. An alternative explanation 
is light curve modulation on time scales of several days, yielding a broad continuous frequency 
spectrum.

The amplitudes of detected pulsation modes span range from as large as 233 mmag for V20 
till 2.5 mmag in V12 where scatter of observations was particularly small (Table \ref{t10}).
V8 and V19 exhibit SX Phe mode pulsation in one mode only, with a harmonic. Low frequency 
modulation in them is weak. In most stars the detected modes span 10-25 c/d frequency range 
oscillations with some harmonics detectable beyond our detection limit of 75 c/d. There is 
also evidence of low frequency modulation of amplitudes reaching 10 mmag and more. 

The strongest evidence of modulation of oscillations is present in V5, as indicated by amplitude
errors over 10 times larger than those resulting from oscillations of comparable strength in other 
stars. We refrain from listing frequencies of several nominal large amplitude oscillations, of 
order $\sim120$ mmag, at frequencies  0.267763, 1.269207, 10.332841, 17.361620, 40.365515 and  
68.385442 c/d, because of their errors exceeding 30 mmag or more. They are either unstable, or 
phase and/or amplitude modulated.

Low frequency oscillations may be due to $\gamma$ Doradus oscillations, or a combination of high 
frequency modes with rotation/orbital modulation. Both the base frequency 1.09 c/d and its harmonic 
are detected in V15. This may indicate rotation modulation, and although opinions in literature differ, 
such a value is consistent with already observed elsewhere (NB 2017, Kurtz et al, 2015).
In three stars, namely V10, V13 and V18, the low amplitude modulation amplitude dominate those of 
high frequency modes. This seems to exclude an origin due to combination of high frequency modes. 
The presence of two such modes in V13 excludes the rotation interpretation, independently on 
whether the true frequency of the stronger mode is coser to 1 or 0 c/d. The remaining explanation 
is: we see $\gamma$ Dor oscillations. Similarly, in Kepler power spectra of 9244992, 6780873, and 
5390069 in NB17 two strong low frequency peaks appear without evidence of candidate combination 
peaks at high frequencies. Because of number, variety and strength of features observed in M10 at 
frequencies below 5 c/d it is hard to avoid the conclusion that while most of them occur due to 
combination of higher frequency modes (in accordance with Kurtz et al. 2015), some of them are 
gravity modes of $\gamma$ Dor type.

In several stars, notably V5, V10, V17, V18 and V19 broad low amplitude bumps still remain in the 
pre-whitened frequency spectrum after removal of all identified discrete frequencies. They 
spread over the typical range of SX Phe base frequencies, sometimes reaching the harmonic region of 
50 c/d. Hence, they seem to be a genuine effect of stellar origin, similar to observed in some 
$\delta$ Scuti stars (Barcel\'o et al. 2017). If so, they may correspond to chaotic instability/modulation 
over a time interval of several base mode cycles. Summarising, our observations reveal low frequency 
oscillations due to either $\gamma$ Dor or combination modes, strong modulation of SX Phe oscillations 
and instability of those modes on time scales as short as days in the extreme.

\clearpage

\begin{table}\label{t10}
\caption{Pulsation frequencies detected in SX Phe stars}
\begin{minipage}[t]{0.50\textwidth}
\begin{tabular}{lrrr}
\hline
	Id & Frequency && Ampl. \\
           & ~~[c/d]~~~~&& [mmag]\\[2mm]
\hline
\multicolumn{4}{c}{V5}\\ 
f$_1$ &   0.623036&  (6) &   49.5(3.7)\\
f$_2$ &   1.011843&  (9) &   34.6(4.3)\\
f$_3$ &   1.424943& (12) &   22.9(3.7)\\
f$_4$ &   1.980009& (10) &   22.3(3.8)\\
f$_5$ &   2.205710&  (6) &   32.0(3.4)\\
2f$_5$ &   4.411421&   - &   16.2(3.1)\\
f$_6$ &   5.809590&  (9) &   24.9(3.0)\\
f$_7$ &  12.180899& (11) &   18.3(3.0)\\
f$_8$ &  17.077723&  (5) &   54.1(4.5)\\
f$_9$ &  17.079976&  (1) &  211.2(4.8)\\
f$_{10}$ &  21.884772& ( 6) &   38.2(3.1)\\
f$_{11}$ &  23.820710& (11) &   17.2(3.2)\\
2f$_9$ &  34.159952&   - &   47.5(3.0)\\
f$_{12}$ &  34.549065& (11) &   17.6(3.0)\\
f$_{13}$ &  35.162365& (13) &   12.9(2.7)\\
f$_{14}$ &  38.964097&  (6) &   31.4(2.9)\\
3f$_9$ &  51.239928&   - &   20.8(3.0)\\
f$_{15}$ &  56.044101& (10) &   18.8(3.0)\\
\multicolumn{4}{c}{V6}\\ 
f$_1$ &   0.474062& (30) &    7.4(1.2)\\
f$_2$ &   0.975011&  (8) &   28.2(1.3)\\
f$_3$ &   1.610948& (13) &   16.3(0.9)\\
f$_4$ &   2.276423& (24) &   10.0(0.9)\\
f$_5$ &   2.372293& (21) &   11.8(1.0)\\
f$_6$ &   2.698967& (15) &   13.8(0.9)\\
f$_7$ &   9.125513& (44) &    4.4(0.7)\\
f$_8$ &  12.368305& (46) &    4.1(0.7)\\
f$_9$ &  13.059566&  (4) &   45.7(0.7)\\
f$_{10}$ &  16.696109&  (3) &   60.5(0.8)\\
f$_{11}$ &  18.087565& (28) &    7.2(0.8)\\
f$_{12}$ &  20.908996& (36) &    5.4(0.7)\\
f$_{13}$ &  23.130960& (54) &    3.9(0.7)\\
f$_{14}$ &  26.117025& (53) &    3.8(0.7)\\
f$_{15}$ &  29.755747& (33) &    5.8(0.7)\\
2f$_{10}$ &  33.392218& -   &    5.9(0.7)\\
\multicolumn{4}{c}{V7}\\ 
f$_1$ &   0.485023& (23) &    6.4(1.4)\\
f$_2$ &   1.681329& (40) &    4.3(0.7)\\
f$_3$ &   2.293711& (43) &    3.6(0.6)\\
f$_4$ &   4.961245& (53) &    3.1(0.6)\\
f$_5$ &  20.787488&  (5) &   34.9(0.7)\\
f$_6$ &  21.348648& (50) &    4.9(0.8)\\
f$_7$ &  21.389382& (25) &    9.8(0.7)\\
f$_8$ &  40.818834& (48) &    3.6(0.7)\\
f$_9$ &  46.769927& (50) &    3.2(0.6)\\
\hline
\end{tabular}
\end{minipage}
\begin{minipage}{0.50\textwidth}
\begin{tabular}{lrrrr}
\hline
	Id & Frequency && Ampl. \\
           & ~~[c/d]~~~~&& [mmag]\\[2mm]
\hline
\multicolumn{4}{c}{V8}\\ 
f$_1$ &   1.744487& (71) &    4.5(0.7)\\
f$_2$ &  19.604362& (7) &    44.7(0.6)\\
2f$_2$ &  39.208725& - &      9.9(0.6)\\
3f$_2$ &  58.813087& - &      4.8(0.6)\\
\multicolumn{4}{c}{V10}\\ 
f$_1$ &   0.642146& (28) &   11.3(1.1)\\
f$_2$ &   1.005681& (33) &   19.1(2.3)\\
2f$_1$ &   1.284291& - &      5.5(1.0)\\
f$_3$ &  14.617339& (111) &   4.0(0.8)\\
f$_4$ &  19.442675& (152) &   4.0(0.8)\\
f$_5$ &  39.761490& (28) &   14.6(0.8)\\
f$_6$ &  42.742695& (87) &    5.1(0.8)\\
f$_7$ &  44.805034& (25) &   18.2(0.8)\\
f$_8$ &  49.849907& (103) &   3.5(0.7)\\
\multicolumn{4}{c}{V12}\\ 
f$_1$ &  22.717670& (41) &    2.9(0.4)\\
f$_2$ &  29.509114& (49) &    2.5(0.4)\\
f$_3$ &  43.782017& (24) &    4.9(0.4)\\
\multicolumn{4}{c}{V13}\\ 
f$_1$ &   1.004305& (37) &   10.5(1.5)\\
f$_2$ &   1.129659& (66) &    5.2(0.8)\\
f$_3$ &   1.360120& (41) &    9.6(0.9)\\
f$_4$ &   9.087800& (86) &    3.3(0.7)\\
f$_5$ &   9.929391& (91) &    3.9(0.8)\\
f$_6$ &  21.471789& (73) &    5.3(0.8)\\
f$_7$ &  22.796205& (65) &    5.8(1.2)\\
f$_8$ &  35.483238& (87) &    4.0(0.8)\\
\multicolumn{4}{c}{V14}\\ 
f$_1$ &   0.643164& (34) &   10.4(1.3)\\
f$_2$ &   0.941986& (54) &    7.1(1.3)\\
f$_3$ &   7.469010& (60) &    6.1(1.1)\\
f$_4$ &  23.138565& (51) &    7.9(1.1)\\
f$_5$ &  26.145759& (29) &   14.1(1.4)\\
f$_6$ &  26.969884& (54) &    6.9(1.2)\\
f$_7$ &  27.048688& (28) &   13.7(1.4)\\
\multicolumn{4}{c}{V15}\\  
f$_1$ &   0.975006& (37) &    4.5(0.7)\\
f$_2$ &   1.089195& (21) &    3.1(0.6)\\
2f$_2$ &   2.178390& - &      3.7(0.6)\\
f$_3$ &  22.816304& (27) &    5.5(0.6)\\
f$_4$ &  23.589314& (49) &    2.9(0.6)\\
f$_5$ &  25.023657& (37) &    3.9(0.6)\\
f$_6$ &  27.784002& (48) &    3.3(0.6)\\
f$_7$ &  28.706828& (8) &    19.5(0.6)\\
f$_8$ &  29.581976& (19) &    8.0(0.6)\\
\hline
\end{tabular}
\end{minipage}\hspace{0.15\textwidth}
\end{table} 

\begin{table}\label{t11}
\caption{Pulsation frequencies detected in SX Phe stars}
\begin{minipage}[t]{0.50\textwidth}
\begin{tabular}{lrrr}
\hline
	Id & Frequency && Ampl. \\
           & ~~[c/d]~~~~&& [mmag]\\[2mm]
\hline
\multicolumn{4}{c}{V17}\\ 
f$_1$ &   0.028952& (20) &    8.6(0.7)\\
f$_2$ &   0.393907& (24) &    5.7(0.6)\\
f$_3$ &   2.038169& (49) &    3.1(0.6)\\
f$_4$ &  27.066525& (4) &    30.8(0.5)\\
f$_5$ &  27.611165& (42) &    3.3(0.6)\\
f$_6$ &  27.786646& (13) &    8.3(0.5)\\
f$_7$ &  52.283249& (25) &    5.5(0.5)\\
2f$_4$ &  54.133050& - &      2.9(0.5)\\
2f$_6$ &  55.573291& - &      2.8(0.5)\\
f$_8$ &  56.065747& (42) &    3.3(0.5)\\
\multicolumn{4}{c}{V18}\\  
f$_1$ &   0.069108& (16) &   10.8(0.6)\\
f$_2$ &   1.398965& (31) &    4.7(0.7)\\
f$_3$ &   1.647285& (38) &    4.8(0.6)\\
f$_4$ &   2.623331& (44) &    4.3(0.8)\\
f$_5$ &  20.542125& (50) &    2.7(0.5)\\
f$_6$ &  23.564537& (14) &    9.6(0.5)\\
\multicolumn{4}{c}{V19}\\ 
f$_1$ &   0.643866& (49) &    2.9(0.6)\\
f$_2$ &   1.001649& (40) &    5.8(0.5)\\
f$_3$ &  22.833647& (6) &    21.1(0.5)\\
f$_4$ &  45.667295& - &       2.8(0.5)\\
\hline
\end{tabular}
\end{minipage}
\begin{minipage}{0.50\textwidth}
\begin{tabular}{lrrr}
\hline
	Id & Frequency && Ampl. \\
           & ~~[c/d]~~~~&& [mmag]\\[2mm]
\hline
\multicolumn{4}{c}{V20}\\ 
f$_1$ &   0.004405& (27) &    7.5(0.7)\\
f$_2$ &   0.121073& (25) &    5.5(0.6)\\
f$_3$ &   2.269694& (45) &    3.1(0.5)\\
f$_4$ &   6.629676& (43) &    3.2(0.5)\\
f$_5$ &  19.112973& (12) &    5.1(0.6)\\
f$_6$ &  19.761623& (1) &   233.2(0.5)\\
f$_7$ &  20.441022& (39) &    3.5(0.5)\\
f$_8$ &  26.356067& (19) &    7.3(0.5)\\
f$_9$ &  37.135945& (39) &    3.5(0.5)\\
2f$_5$ &  38.225946& - &      5.5(0.6)\\
f$_{10}$ &  38.961059& (33) &    4.2(0.6)\\
2f$_6$ &  39.523246& - &     75.2(0.6)\\
f$_{11}$ &  40.367636& (42) &    3.0(0.5)\\
f$_{12}$ &  46.114884& (30) &    4.6(0.5)\\
f$_{13}$ &  47.232920& (54) &    2.6(0.5)\\
f$_{14}$ &  58.991013& (51) &    2.6(0.5)\\
3f$_6$ &  59.284869& - &     22.8(0.5)\\
4f$_6$ &  79.046492& - &     11.3(0.5)\\
5f$_6$ &  98.808115& - &      5.2(0.5)\\
&&&\\
&&&\\
&&&\\
\hline
\end{tabular}
\end{minipage}\hspace{0.15\textwidth}
\end{table}

\begin{table}[H]
\footnotesize
 \begin{center}
 \caption{\footnotesize Basic data of the variables identifed in the M10 field\strut}
          \label{tab:CASE}
 \begin{tabular}{|l|c|c|c|c|c|c|c|c|c|}
  \hline
 ID & RA$_{\mathrm J2000}$ & DEC$_{\mathrm J2000}$ & $V$ & $B-V$ & $\Delta_V$ &Period & Type$^a$ & M$^b$\\
    & [deg] & [deg]  &[mag]& [mag] & [mag] & [d] &             &\\
  \hline
 V1  &  254.29214  &  -4.09336  &  11.83  &  1.52  &  0.56  &  70.878903  &  SR  &  Y \\
 V2  &  254.29891  &  -4.06658  &  12.05  &  0.96  &  1.18  &  19.470995  &  W Vir  &  Y \\
 V3  &  254.23315  &  -4.07123  &  12.75  &  0.87  &  0.47  &  7.872181  &  W Vir  &  Y  \\
 V4  &  254.31004  &  -4.18284  &  14.74  &  0.53  &   --   &     --     &  $const$  &  Y  \\
 V5  &  254.28580  &  -4.10453  &  16.97  &  0.44  &  0.55  &  0.058543  &  SX  &  Y \\
 V6  &  254.29457  &  -4.09261  &  16.69  &  0.51  &  0.20  &  0.059909  &  SX  &  Y  \\
 V7  &  254.29321  &  -4.11758  &  17.48  &  0.50  &  0.09  &  0.048112  &  SX  &  Y \\
 V8  &  254.28491  &  -4.08576  &  16.99  &  0.46  &  0.11  &  0.051009  &  SX  &  Y* \\
 V9  &  254.29404  &  -4.09771  &  17.20  &  0.68  &  0.27  &  0.051301  &  SX?  &  Y* \\
V10  &  254.28512  &  -4.11522  &  17.39  &  0.63  &  0.08  &  0.022319  &  SX  &  Y     \\
V11  &  254.29507  &  -4.09886  &  16.91  &  0.80  &  0.13  &  0.047957  &  DS  &  N     \\
V12  &  254.27108  &  -4.10487  &  17.18  &  0.43  &  0.04  &  0.022823  &  SX  &  Y     \\
V13  &  254.28890  &  -4.06591  &  16.89  &  0.48  &  0.06  &  0.036174  &  SX  &  Y     \\
V14  &  254.28831  &  -4.10149  &  17.32  &  0.46  &  0.09  &  0.038198  &  SX  &  Y     \\
V15  &  254.30533  &  -4.09694  &  17.44  &  0.53  &  0.07  &  0.034835  &  SX  &  Y     \\
V16  &  254.27802  &  -4.14490  &  16.76  &  0.92  &  0.03  &  0.357809  &  $susp^c$ &  Y  \\
V17  &  254.27298  &  -4.12981  &  17.24  &  0.52  &  0.09  &  0.036944  &  SX  &  Y     \\
V18  &  254.33431  &  -4.08116  &  17.51  &  0.50  &  0.05  &  0.042435  &  SX  &  Y     \\
V19  &  254.41109  &  -4.14928  &  17.61  &  0.53  &  0.06  &  0.043795  &  SX  &  Y     \\
V20  &  254.26237  &  -4.06683  &  16.97  &  0.44  &  0.51  &  0.050603  &  SX  &  Y     \\
V21  &  254.31559  &  -4.10557  &  19.31  &  0.86  &  0.21  &  0.244976  &  EW  &  Y     \\
V22  &  254.28467  &  -4.03882  &  14.62  &  0.47  &  0.34  &  0.404604  &  RRc  &  Y    \\
V23  &  254.26299  &  -4.13106  &  17.58  &  0.53  &  0.06  &  1.446583  &  $sin$  &  Y    \\
V24  &  254.28146  &  -4.09510  &  13.95  &  0.76  &  0.35  &  2.307458  &  BL Her  &  Y  \\
V25  &  254.27427  &  -4.06279  &  17.41  &  -0.02  &  0.07  &  4.457001  &  SR  &  Y    \\
V26  &  254.30499  &  -4.06993  &  16.36  &  1.04  &  0.29  &  21.784707  &  $sin$  &  Y   \\
V27  &  254.31281  &  -4.09790  &  11.91  &  1.89  &  0.28  &  21.004000  &  SR  &  Y    \\
V28  &  254.29483  &  -4.07884  &  11.90  &  1.83  &  0.25  &  60.483833  &  SR  &  Y    \\
V29  &  254.36407  &  -4.02354  &  11.87  &  1.76  &  0.58  &  68.388291  &  SR  &  Y    \\
V30  &  254.28242  &  -4.10166  &  12.43  &  1.42  &  0.24  &  71.667981  &  SR  &  Y    \\
V31  &  254.25264  &  -4.07014  &  15.92  &  0.09  &  0.07  &  0.205066  &  $sin,susp$  &  Y\\
V32  &  254.36188  &  -4.07538  &  17.93  &  0.73  &  0.04  &  0.848041  &  $sin,susp$  &  Y  \\
V33  &  254.34408  &  -4.05701  &  17.56  &  0.00  &  0.09  &  0.933530  &  $sin,susp$  &  Y  \\
V34  &  254.29489  &  -4.14371  &  17.01  &  0.91  &  0.16  &  3.339100  &  $sin^d$  &  Y  \\
 N1  &  254.20253  &  -3.97676  &  19.64  &  0.43  &  0.34  &  0.063696  &  DS$^e$  &  N \\
 N2  &  254.27678  &  -3.92101  &  20.10  &  1.11  &  0.35  &  0.268554  &  EW  &  N \\
 N3  &  254.34280  &  -4.08330  &  16.54  &  0.48  &  0.42  &  0.294384  &  RRc  &  N  \\
 N4  &  254.27512  &  -4.23526  &  16.74  &  1.64  &  0.08  &  2.111434  & $sin$ &  N  \\
 N5  &  254.27756  &  -4.21678  &  18.00  &  1.16  &  0.24  &  6.766509  & $sin$ &  N \\
 N6  &  254.27708  &  -4.01820  &  22.10  &  ---  &  0.30  &  0.041981  &  EA/EB  &  N \\
  \hline
 \end{tabular}
\end{center}
{\footnotesize
$^a$ EA - detached eclipsing binary, EB - type $\beta$~Lyr eclipsing binary, EW - contact binary; 
SX, DS, RRc, W Vir and BL Her - pulsators of SX Phe, $\delta$~Sct, RR Lyr c, W Vir and BL Her type,   
$sin$ - sinusoidal light curve of unknown origin, $susp$ - suspected variable, $const$ - no 
variability detected.\\
$^b$ Membership status according to N17: Y - member or likely member (starred: discrepant CASE and 
{\it Gaia} proper motions; see Appendix), N - field object.\\
$^c$ Variability detected in 2015 only.~~ $^d$ M10-VLA1 (S18).~~$^e$ V3 of von Braun et al. (2002).}
\end{table}

\clearpage

\begin{figure}[H]
    \centerline{\includegraphics[width=0.99\textwidth,
                bb = 130 207 508 585 clip]{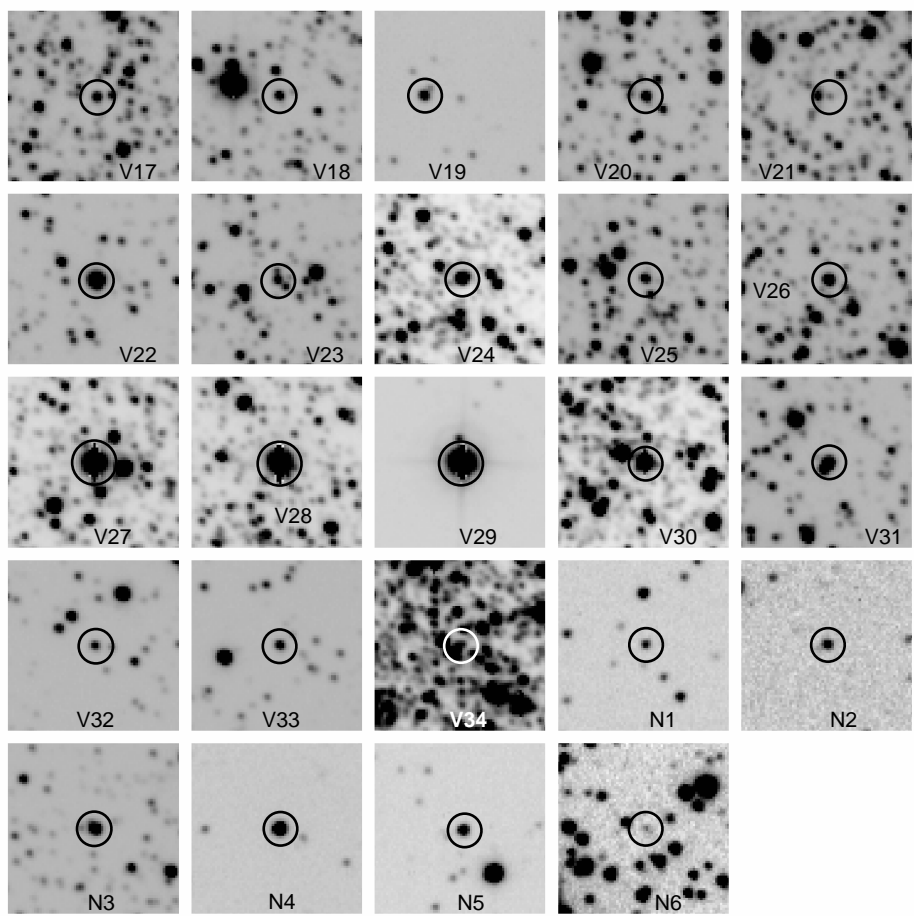}}
    \caption{Finding charts for the new variables whose light curves are shown
     in Figs. \ref{fig:CASE1} and \ref{fig:CASE2}. 
     Each chart is 30\arcs on a side. North is up and East to the left.
     \label{fig:maps}}
\end{figure}

 \begin{figure}[H]
    \centerline{\includegraphics[width=0.95\textwidth,
                bb = 42 25 528 764 clip]{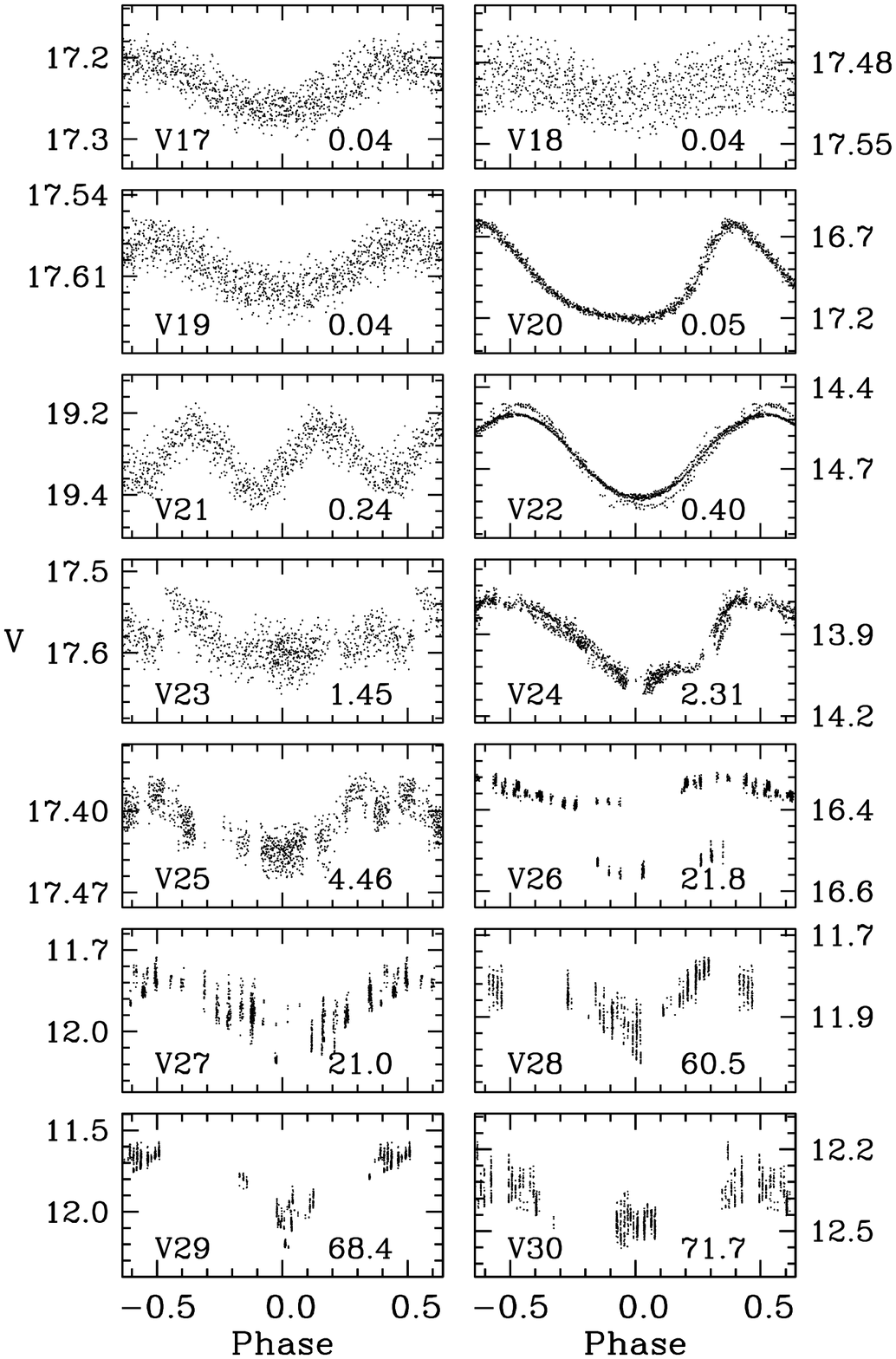}}
    \caption{Phased $V$-band light curves for the newly discovered variables. 
    Panel labels give star ID and period in days.
     \label{fig:CASE1}}
 \end{figure}
 
 \begin{figure}[H]
    \centerline{\includegraphics[width=0.95\textwidth,
        bb = 42 226 528 764, clip]{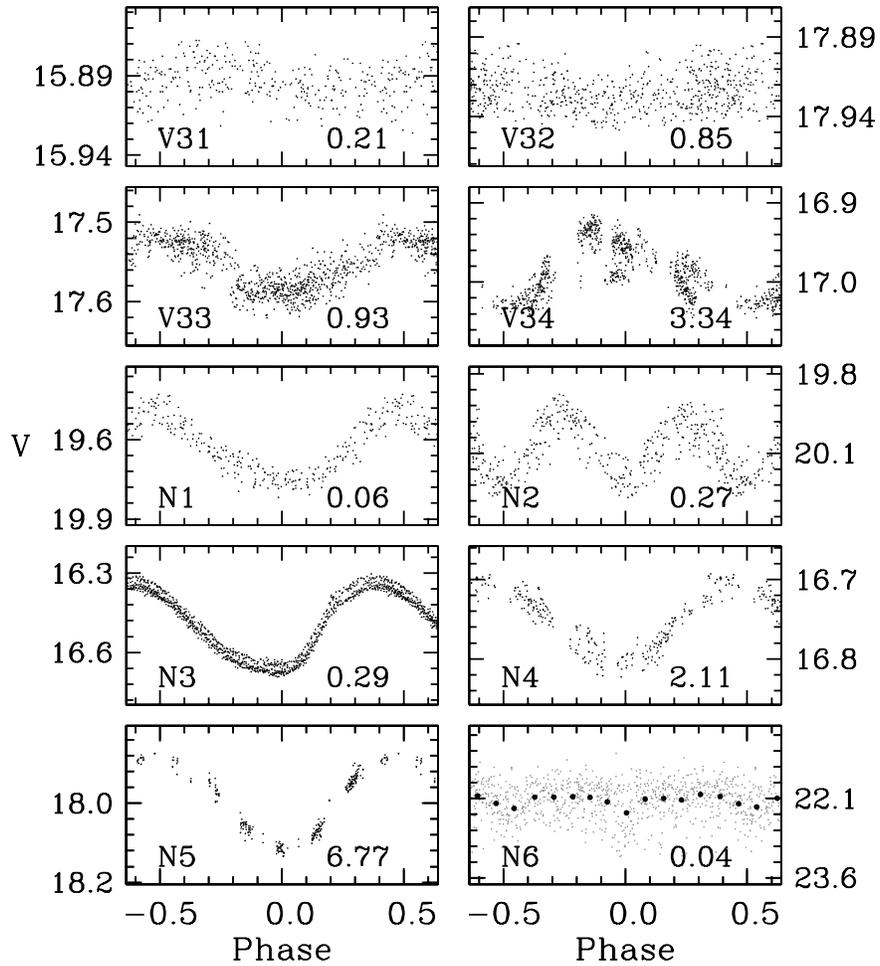}}
   \caption{Continuation of Fig. \ref{fig:CASE1} Phase-binned data for N6 are shown with 
    heavy black points.
    \label{fig:CASE2}}
\end{figure}

\end{document}